\documentclass[aps,prl,twocolumn,superscriptaddress]{revtex4-1}
\usepackage[dvips]{color,epsfig}
\usepackage{amsmath}
\usepackage{graphicx}
\usepackage{amssymb}
\usepackage{amsthm, amscd}
\usepackage{amsfonts}
\usepackage{subfigure}

\newcommand{\bscco}{Bi$_2$Sr$_2$CaCu$_2$O$_{8+\delta}$}
\newcommand{\ccoc}{Ca$_{2-x}$Na$_x$CuO$_2$Cl$_2$}
\newcommand{\lsco}{La$_{2-x}$Sr$_x$CuO$_4$}
\newcommand{\lbco}{La$_{2-x}$Ba$_x$CuO$_4$}
\newcommand{\ybco}{Y$_{2-x}$Ba$_x$CuO$_4$}

\begin{document}

\title{Nodal-antinodal dichotomy from pairing disorder in d-wave superconductors}

\author{Dimitrios Galanakis}
\affiliation{Department of Physics and Astronomy, Louisiana State University, Baton Rouge, Louisiana, 70803, USA}
\author{Stefanos Papanikolaou}
\affiliation{Department of Physics, Cornell University, Ithaca, 14853-8150, New York, USA}

\date{\today}

\begin{abstract}
We study the basic features of the local density of states~(LDOS) observed in STM experiments on high-T$_c$ d-wave superconductors in the context of a minimal model of a d-wave superconductor which has {\it weakly} modulated off-diagonal disorder. We show that the low and high energy features of the LDOS are consistent with the observed experimental patterns and in particular, the anisotropic local domain features at high energies. At low energies, we obtain not only the scattering peaks predicted by the octet model~\cite{kohsaka07}, but also weak features that should be experimentally accessible. Finally, we show that the emerging features of the LDOS lose their correspondence with the features of the imposed disorder, as its complexity increases spatially.  
\end{abstract}

\pacs{PACS numbers:}
\maketitle
In recent years, the effect of underlying inhomogeneities in
superconductors has been studied intensively. In the context of
high temperature superconductors (HTSC),  checkerboard local
density of states (LDOS) oscillations and strong nanoscale gap
inhomogeneity have been observed in scanning tunneling
spectroscopy (STS) experiments \cite{fischerreview}, and signals
in dynamical susceptibility measured by neutron scattering have
been interpreted as stripe-like nanoscale modulations of charge and
spin degrees of freedom \cite{Tranquadareview}. Even though it is still
not clear whether these modulations are intrinsic or
driven entirely by disorder, it seems plausible that the magnitude of $T_c$ might be 
strongly related to the very existence of inhomogeneity on the coherence length scale~\cite{martin05, arrigoni03}, and
therefore a deeper understanding of the source of inhomogeneities is important. In the cases of the compounds \lbco and \lsco (with Nd or Eu co-doping), inhomogeneities take the form of static long-range stripe-like spin and charge modulations~\cite{lsco-stripe,lnsco-stripe,pnas-emery,neutrons-lbco}.


Tunneling spectroscopy has been used to probe states in different regions of momentum and energy, by the help of the Fourier transform scanning tunneling spectroscopy (FT-STS) and the high energy features of the LDOS. At low energies, in the d-wave superconducting state near optimal doping, quasiparticle interference (QPI) observed by FT-STS is dominated by peaks at well-defined wavevectors ${\bf q}_i$ which are consistent with a simple model of Bogoliubov quasiparticles, called the octet model~\cite{hoffman02,stm-davis-05}. Even though the qualitative features of the octet model are experimentally robust, a quantitative understanding of the amplitude, location and width of the peaks is not straightforward to obtain and depends rather sensitively on the nature of the scattering medium~\cite{andersen08}. A point-like scatterer, for example, in an otherwise homogeneous dSC leads to a landscape of some spot-like and some arc-like dispersive features close to ${\bf q}_i$ in the FT-STS images~\cite{capriotti03}, whereas experimental data appear mostly spot-like. At high energies, local unidirectionality has been observed in domains with size approximately $5a_0$, close to the superconductor's coherence length and the domains typically alternate in orientation. This behavior is usually attributed to a disordered charge-density wave (CDW), with some success on demonstrating the emergence of the LDOS patterns from the underlying CDW order.~\cite{vojta08}.

In Refs.~\cite{vojta00,vojta00-b} it was shown that generic off-diagonal disorder (coupling of the nodal d-wave quasiparticles to an s-wave order parameter $\Delta_s$) is a relevant operator that leads to a first-order transition in a system with tetragonal-orthorhombic symmetry breaking. Therefore, off-diagonal disorder presents the dominating effect for superconductors with tendency towards orthorhombic symmetry breaking. However, when the orthorhombic symmetry is not broken (due to frustration effects), it is expected that modulated forms of the expected disorder should be present, with no effect at large length scales. In this paper we present a an explicit model with off-diagonal, modulated disorder, which unifies the basic features of the STM observations into a consistent framework.

In this model, itinerant fermions are coupled to weak, off-diagonal disorder, modulated in well-defined domains. Such a state is characterized by a density wave or modulated nematic order. We motivate the existence of such disorder on the basis of: i) a phenomenological theory of competing s and d-order parameters ii) recent mean-field studies on more detailed states, supporting a modulation of the off-diagonal disorder~\cite{vojta08} . First, we show by using first order perturbation theory, that the spectrum at low energies and the position of the nodes remains unchanged, an expected result since at long length scales the disorder averages to zero. Secondly, we find that at low energies the Fourier-maps of the LDOS have peaks that are dispersing according to the octet-model scattering~\cite{andersen08}, given that $\Delta^{(s)}/\Delta_d\ll 1$ with the addition of weak static peaks, bearing strong similarities to the STM experiments' observations~\cite{kohsaka07}. In addition there are static peaks close to $(\pm 3\pi/5,\pm\pi/5)$, but these peaks are very weak (since $\Delta^{(s)}$ is very small) and if spatial disorder exists, span a large region around these wavevectors. Thirdly, we show that at high energies the LDOS acquires a domain structure that is locally anisotropic and highly similar to the observed STM patterns. The order parameter~\cite{kivelson03} we use to identify the anisotropy, shows that by increasing the disorder amplitude, the anisotropy becomes noticeable at higher energies. Such anisotropy is not evident in a system with disorder that respects the d-wave symmetry. Finally, we show that disorder with irregular features, coming, for example, from an off-critical system, has \emph{no direct correspondence} to the LDOS patterns, and at high energies there are no local anisotropic features, due to the complexity of the high energy spectrum and induced bound-states~\cite{andersen06}.
\begin{figure}
\subfigure[]{\includegraphics[height=0.16\textwidth]{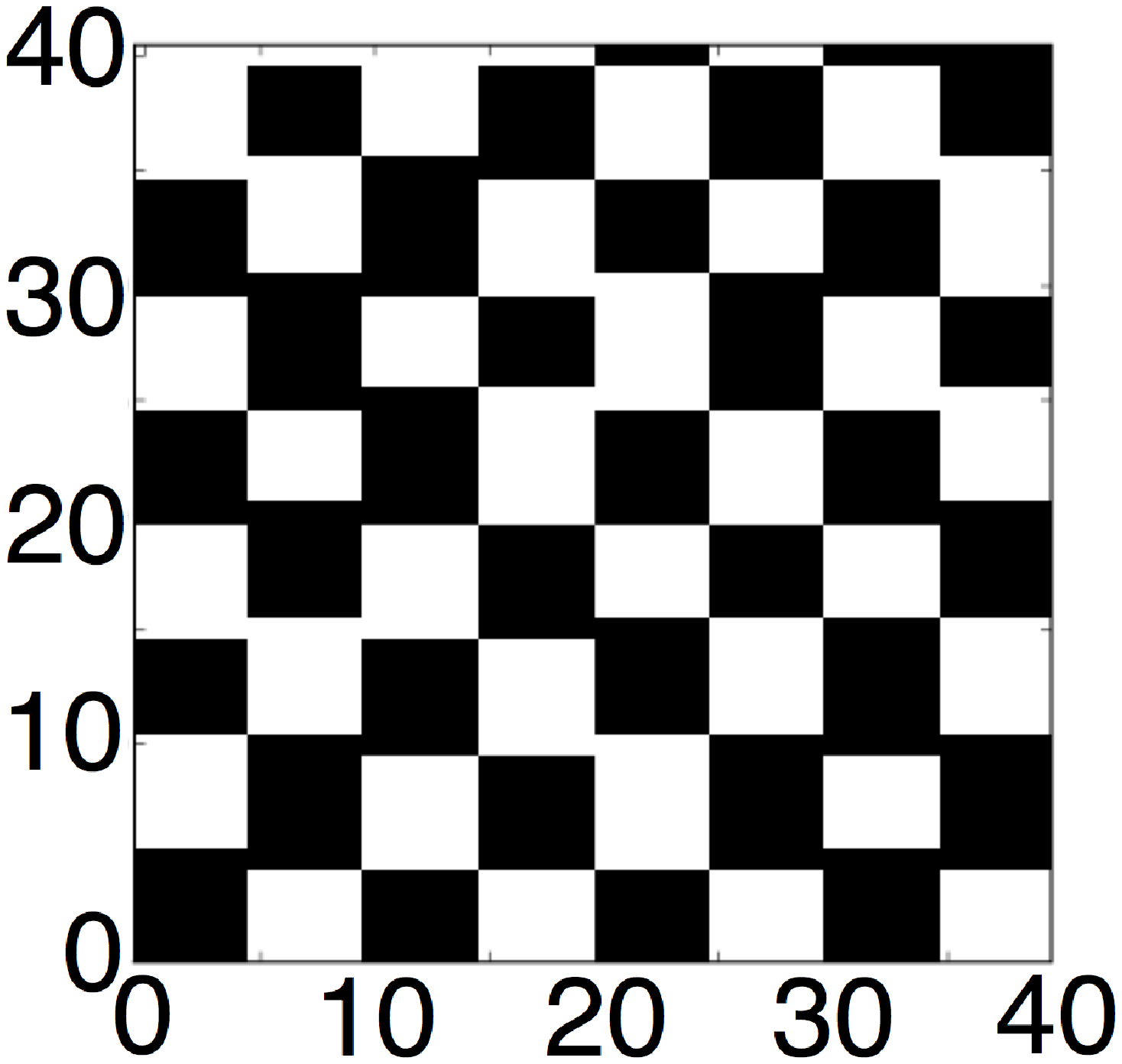}\label{fig:1sub1}}
\subfigure[]{\includegraphics[height=0.18\textwidth]{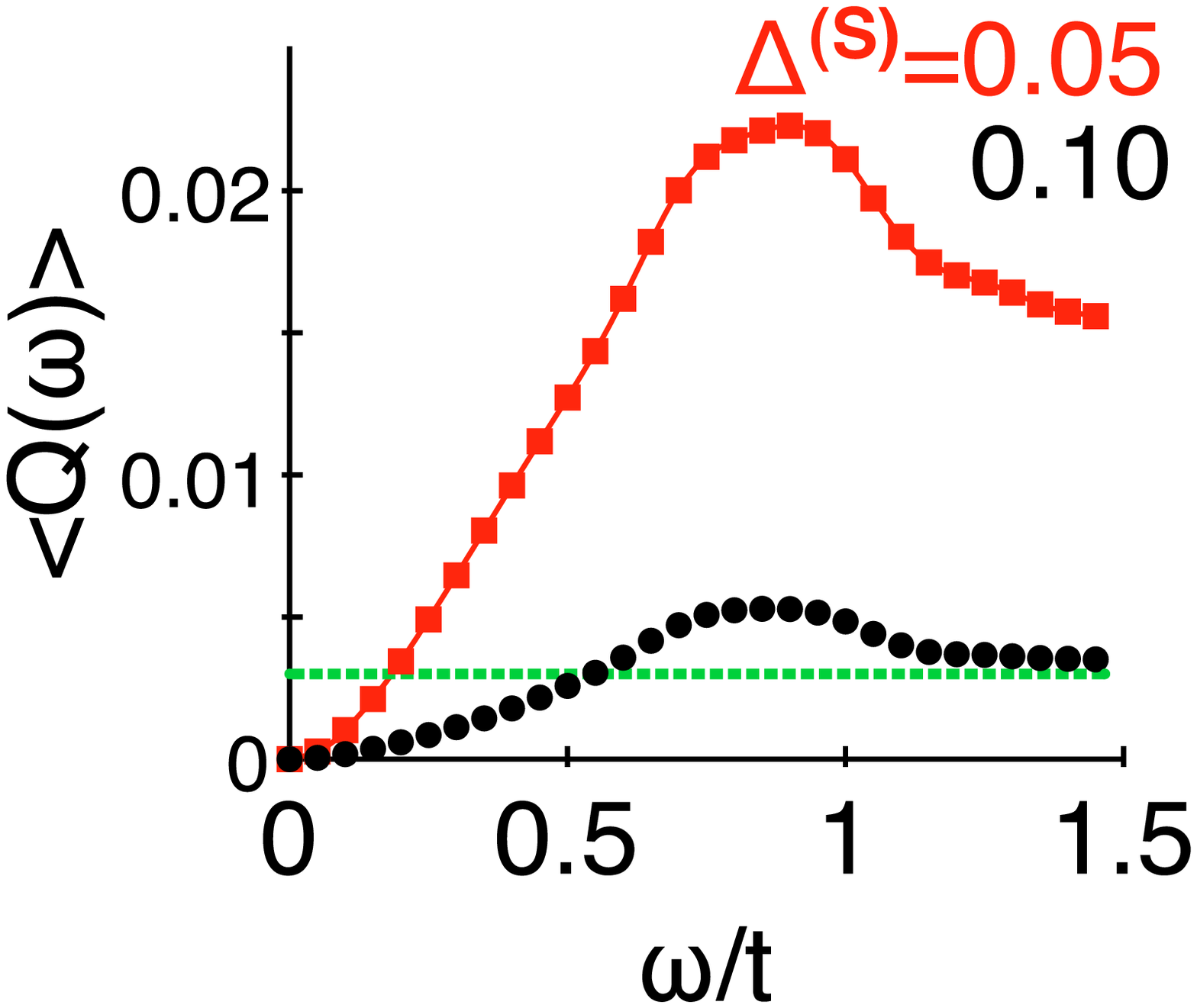}\label{fig:1sub2}}
\caption{{\bf Off diagonal disorder and the anisotropy} (a) A typical form of the employed off-diagonal disorder for $L=40$. The off-diagonal component has a weakly disordered modulation wavelength which is assumed to be close to the superconducting coherence length. In (b), the order parameter of the anisotropy, as defined in the text, shows a strong a peak at 0.8t, and decays slowly at higher energies. The decay is associated with the fact that the domains become internally homogeneous at very high energies, even though there is a remaining anisotropy. Given an experimental resolution, it is clear that decreasing the magnitude of $\Delta_s$ makes the anisotropy observable at higher energies.}
\label{fig:1}
\end{figure}

To motivate the existence of modulated off-diagonal disorder,  we consider a system which potentially can support both s- and d- wave order parameters $\Psi_s=|\Psi_s|e^{{\bf i}\phi_s}$ and  $\Psi_d=|\Psi_d|e^{{\bf i}\phi_d}$. The Ginsburg-Landau functional takes the form
\begin{eqnarray}
F[\Psi_d,\Psi_s]&=&a_d|\Psi_d|^2+\frac{b_d}{2}|\Psi_d|^4+K_d|\nabla\Psi_d|^2\nonumber\\
&+&a_s|\Psi_s|^2+\frac{b_s}{2}|\Psi_s|^4+K_s|\nabla\Psi_s|^2\nonumber\\
&+&a_{sd}(\Psi_{s}^*\Psi_d+c.c.)+b_{sd}|\Psi_s|^2|\Psi_d|^2,
\label{eq:GL}
\end{eqnarray}
where $a_{sd}$ is the coupling between the two order parameters and $b_{sd}>0$ is a repulsive interaction that suppresses all s-d mixing in the absence of the bilinear $a_{sd}$ term. Furthermore it is assumed that $a_d<a_s<0$, $b_d,b_s>0$, $a_{sd}$. The energy of phase boundaries is $F_B=-J_s|\Psi_{s}^1||\Psi_{s}^2|\cos(\phi_{s}^1-\phi_{s}^2)-J_d|\Psi_{d}^1||\Psi_{d}^2|\cos(\phi_{d}^1-\phi_{d}^2)$, where $1,2$ label neighboring grains. In the limit where $J_s<J_d$, $b_{sd}|\Psi_d|^2>J_s$ and $a_s\ll a_d$, which is a natural limit for the cuprates, minimizing the free energy favors {\it generically} the proliferation of phase domain walls of the s-order parameter where the $b_{sd}$ repulsion energy is suppressed~\cite{galanakis2}, and the s-order parameter becomes zero on the boundary. Since the grain boundaries are defined within the coherence length scale, the off-diagonal order parameter $\Delta_s$ should be modulated over distances larger than the coherence length. Additional motivations for the consideration of a modulated mixed SC order parameter come from microscopically motivated mean-field states~\cite{vojta08}, where the coexistence of modulated mixed order parameters is necessary.

We consider a model of lattice itinerant fermions, in which the Hamiltonian takes the form,
\begin{eqnarray}
H_{\rm F}&=&-t\sum_{\left<{\bf ij}\right>\sigma}(c^\dagger_{{\bf i}\sigma}c_{{\bf j}\sigma}+h.c.) +\sum_{{\bf i},\alpha}(\Delta_{{\bf i}\alpha}
c_{{\bf i}\uparrow}c_{{\bf i}+\alpha\downarrow}+h.c.)\nonumber\\
&&+ \sum_{{\bf i}}(\Delta^{(s)}_{\bf i}c_{{\bf i}\uparrow}c_{{\bf i}\downarrow}+h.c.), 
\label{eq:hamfermi}
\end{eqnarray} 
\noindent
where $c_{{\bf i}\sigma}$ are fermionic operators, 
and
$\Delta_{{\bf i}\alpha}$=$\beta$$|\Delta_{{\bf i}}|$$e^{i\phi_{\bf i}}$
are complex numbers for the d-wave SC order parameter
defined now at the links (${\bf i}$,${\bf i}$+$\alpha$)($\alpha$=unit vector
along the $x$ or $y$ directions; $\beta$=1 (-1) for $\alpha$ along $x$ ($y$)). 
The order parameter $\Delta^{(s)}_i$ is chosen to be modulated in space and take
the values $\pm|\Delta^{(s)}|$ in short - range domains~\cite{footnote-shortrange}\ref{fig:1sub1}. 

For $\Delta^{(s)}\rightarrow0$ this model can be studied in perturbation theory. If we assume that $\Delta^{(s)}(x,y)=\sin(\phi_q{\bf r})$, the spectral function at first-order takes the form,
\begin{eqnarray}
A(\omega,{\bf q}) & = & \frac{1}{2}G_{1\uparrow}({\bf k},{\bf k},-\omega)+\frac{1}{2}G_{1\downarrow}({\bf k},{\bf k},\omega)\nonumber\\
 & = & \Delta_{s}s_{{\bf q}}\frac{1}{N}\sum_{{\bf k}}\frac{E_{{\bf k}}^{2}\sin\left(\phi_{{\bf k}+{\bf q}}+\phi_{{\bf k}}\right)}{E_{{\bf k}}^{2}-E_{{\bf k}+{\bf q}}^{2}}\frac{1}{\omega^{2}-E_{{\bf k}}^{2}},
\end{eqnarray}
which has poles at the same locations as the unperturbed system. This result shows that for weak disorder amplitudes, the low energy properties of the superconductor and the positions of the nodal points are unaffected, as it is a well known experimental fact for the cuprates~\cite{damascelli03}. 

In the following, we go beyond perturbation theory and we study both low and high energy features of the model by solving it in a self-consistent mean-field manner~\cite{atkinson00}, using exact diagonalization on systems of size $L\times L$. The relevant parameters to our calculation satisfy the hierarchy $\Delta^{(s)}\ll \Delta^{(d)}\ll t$,  with $t\sim 400meV$ and $\Delta^{d}\sim60-100meV$, which we believe to be consistent with the experiments on cuprates. Equation \ref{eq:hamfermi} can be diagonalized by using a Bogoliubov transformation. The corresponding Bogoliubov - de Gennes equations are solved iteratively, for fixed $\Delta^{(s)}$ fixed until a self consistent solution is found for the d-wave order parameter $\Delta_{\bf{i\alpha}}=g\left\langle c_{\bf{i+\alpha}\downarrow}^{\dagger}c_{\bf{i}\uparrow}^{\dagger}\right\rangle $,
\begin{eqnarray}
\left(\begin{tabular}{r r}
$\hat \xi_{\uparrow}$ & $\hat\Delta$ \\ $\hat\Delta^* $ & $-\hat\xi_{\downarrow}^*$ \\
\end{tabular}\right) 
\left( \begin{tabular}{r} $u_n$ \\ $v_n$ \end{tabular}\right) 
= E_n
\left(\begin{tabular}{r} $u_n$ \\ $v_n$ \end{tabular}\right) \label{eq:matrix_eq_uv}
\end{eqnarray}
The mean-field parameters are updated after each iteration using the equation
\begin{eqnarray}
\Delta_{\alpha i} =g\sum_{n} (u_{n,i} v^{*}_{n,i+\alpha} + u_{n,i+\delta}v^{*}_{n,i})\tanh \left(\frac{E_n}{2T}\right)
\end{eqnarray}
where the nearest neighbor interaction $g<0$ has to be attractive in order for the d-wave superconductor to be stable. The coupling $g$ was chosen $g\equiv g_0=-2.5t$ so that the self-consistent average d-wave gap $\Delta^{(d)}$ is $0.26t$. The amplitude of the $s$-wave disorder is chosen as $0.05t$, unless stated otherwise. 
\begin{figure}
\includegraphics[height=0.17\textwidth]{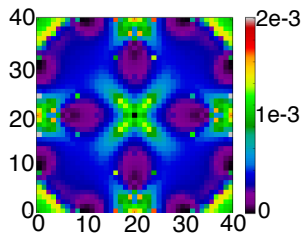}
\centering
\caption{{\bf LDOS at Low energies}. The typical form of the LDOS at a low energy $\omega=0.3\Delta_d$ is shown, for the pure $5a_0$ density wave model, and where a potential impurity is added, in order to compare with the results of Ref.~\cite{wang03}. The amplitude of the Fourier transform of the patterns, shows a number of peaks that are not dispersing with the energy change, but their amplitude is weak and close to the experimentally resolved $q_2$ wavevector~\cite{kohsaka07}. The origin of the peaks is associated with the allowed scattering wavevectors for a d-wave superconductor in the presence of a density wave, similar with Ref.~\cite{wang03}. In that case, the wavevector is $(\pm\pi/4,0)$ and $(0,\pm\pi/4)$, here it is $(\pm\pi/5,\pm3\pi/5)$ and $(\pm3\pi/5,\pm\pi/5)$, very close to what has been labelled as $q_2$.}
\label{fig:2sub1}
\end{figure}

The spectral function of the model (cf. Fig.~\ref{fig:1sub1}) at low energies is similar to the unperturbed d-wave SC, where the low-energy quasiparticles extend along the unperturbed fermi-surface. In order to study the features of the LDOS at low energies in comparison to the well-studied unperturbed case~\cite{wang03, andersen08}, we consider an impurity potential (a change in the chemical potential) at a single site with {\it small} strength $U=0.05t$ in order not to induce any Friedel oscillations. As shown in Fig.~\ref{fig:2sub1}, the LDOS contains similar peaks as the unperturbed system~\cite{andersen08}. In addition, there are more {\it non-dispersing} peaks at ${\bf Q}=(\pm\pi/5,\pm3\pi/5)$ and $(\pm3\pi/5,\pm\pi/5)$, which vanish at larger energies, similar to the experimental observations~\cite{kohsaka07}. These peaks signify the emergence of the locally anisotropic density wave, which at low energies has very weak features. In the presence of spatial disorder, like for example in the model of Fig.~1, these peaks are in close proximity to the denoted wavevectors ${\bf Q}$. Since these peaks are static, they might be present in the region around the observed wave-vector $q_2$~\cite{kohsaka07}. The presence of these disorder peaks could validate the present model.

\begin{figure}
\subfigure[]{\includegraphics[height=0.17\textwidth]{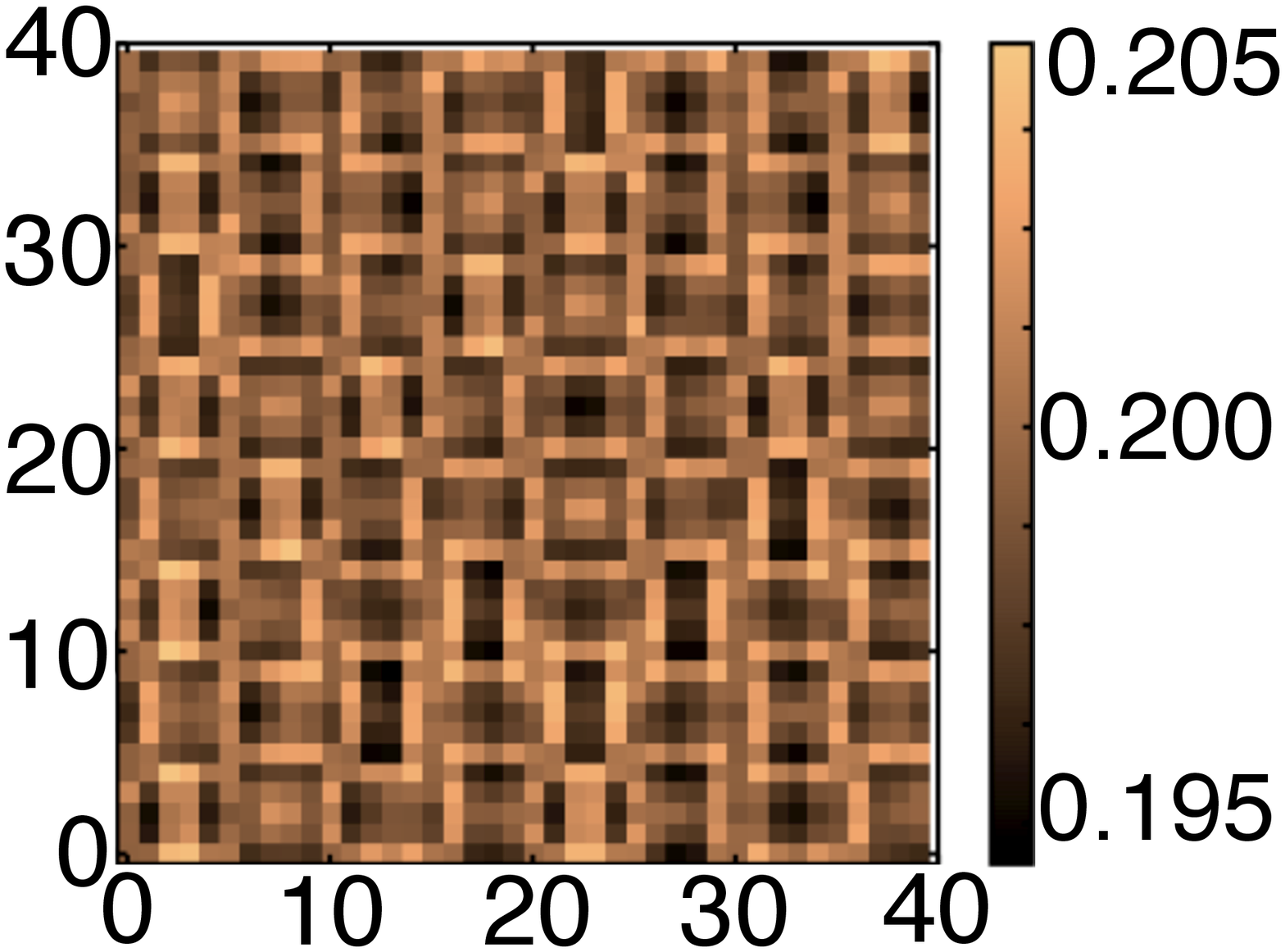}\label{fig:3sub1}}
\subfigure[]{\includegraphics[height=0.17\textwidth]{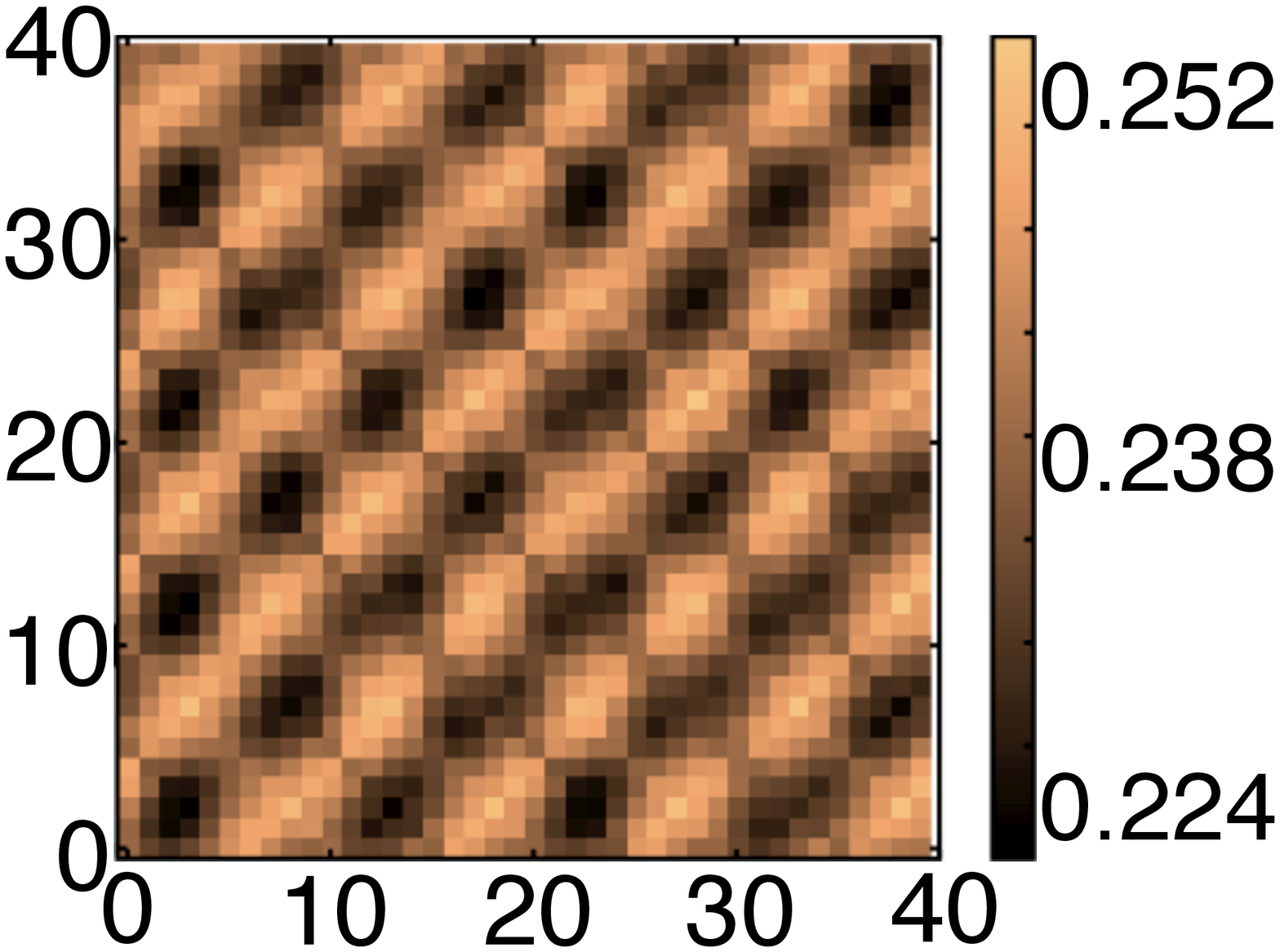}\label{fig:3sub2}}
\caption{{\bf LDOS at high energies}. The form of the LDOS at $\omega=0.7t=2.0\Delta_d$ is shown in (a). The pattern that emerges has strong anisotropic features, similar to the experimentally observed ones. Notice the bond-based structure and the local anisotropy that forms in this case. Such features are robust under weak disorder. In (b), the form of the LDOS when the disorder respects the d-wave symmetry, has no evident anisotropic features at $\omega=t$. Only a natural modulation of the density of states is visible, imposed by the coupling modulation.}
\end{figure}

The weak disorder we described is adequate to induce local anisotropic features at high energies.  As shown in Fig.~\ref{fig:3sub1}, the LDOS shows strong anisotropic features and internal domain structure that resembles the experimental observations. The large intensity near the domain walls on line-like structures and the similarity to large parts of the observed STM maps, signify that our toy description captures important experimental facts. Moreover, we define a local order parameter of the anisotropy~\cite{kivelson03}, applied on the LDOS $F(r)$
\begin{eqnarray}
Q(r)=((\partial_{x}^2-\partial_{y}^2)F(r))^2 + 4((\partial_x\partial_y)F(r))^2
\end{eqnarray}
This order parameter, averaged spatially and over several disorder configurations, distinguishes between anisotropic fluctuations of the LDOS. As shown in Fig.~\ref{fig:1sub2}, the anisotropy increases monotonically until E$_{th}\simeq0.8t$ and then decays slowly at higher energies. The energy threshold where the anisotropy is maximally visible, is almost {\it independent} of the amplitude of the disorder, but dependent on the d-wave gap scale (E$_{th}\sim2\Delta_d)$. Assuming that the experimental resolution allows for a low threshold on the identification of the order parameter (horizontal line in Fig.~\ref{fig:1sub2}) indicates that the smaller the amplitude of the off-diagonal disorder, the higher the energy where the locally anisotropic features of the domains become visible. This observation signifies that the energy scale $(\sim 2\Delta_d)$ where the high-energy domains in STM experiments become visible can be much larger than the actual energy scale which leads to their formation, which it is just $0.3\Delta_d$.  

Another form of possible superconducting disorder is a pair-density wave (PDW) such as the one proposed in association to experiments in \lbco~\cite{berg09} and \bscco~\cite{andersen07}. We investigated the behavior of LDOS in the presence of such a PDW, by setting $\Delta^{(s)}=0$ and modulating the coupling $g=g_0+\delta g$, where $\delta g$ follows the pattern of Fig.~\ref{fig:1sub1} and has a maximum $|\delta g|_{\rm max}=0.05t$. As seen in Fig.~\ref{fig:3sub2}, there is no evidence of local anisotropy due to the presence of such disorder. 

\begin{figure}
\subfigure[]{\includegraphics[height=0.18\textwidth]{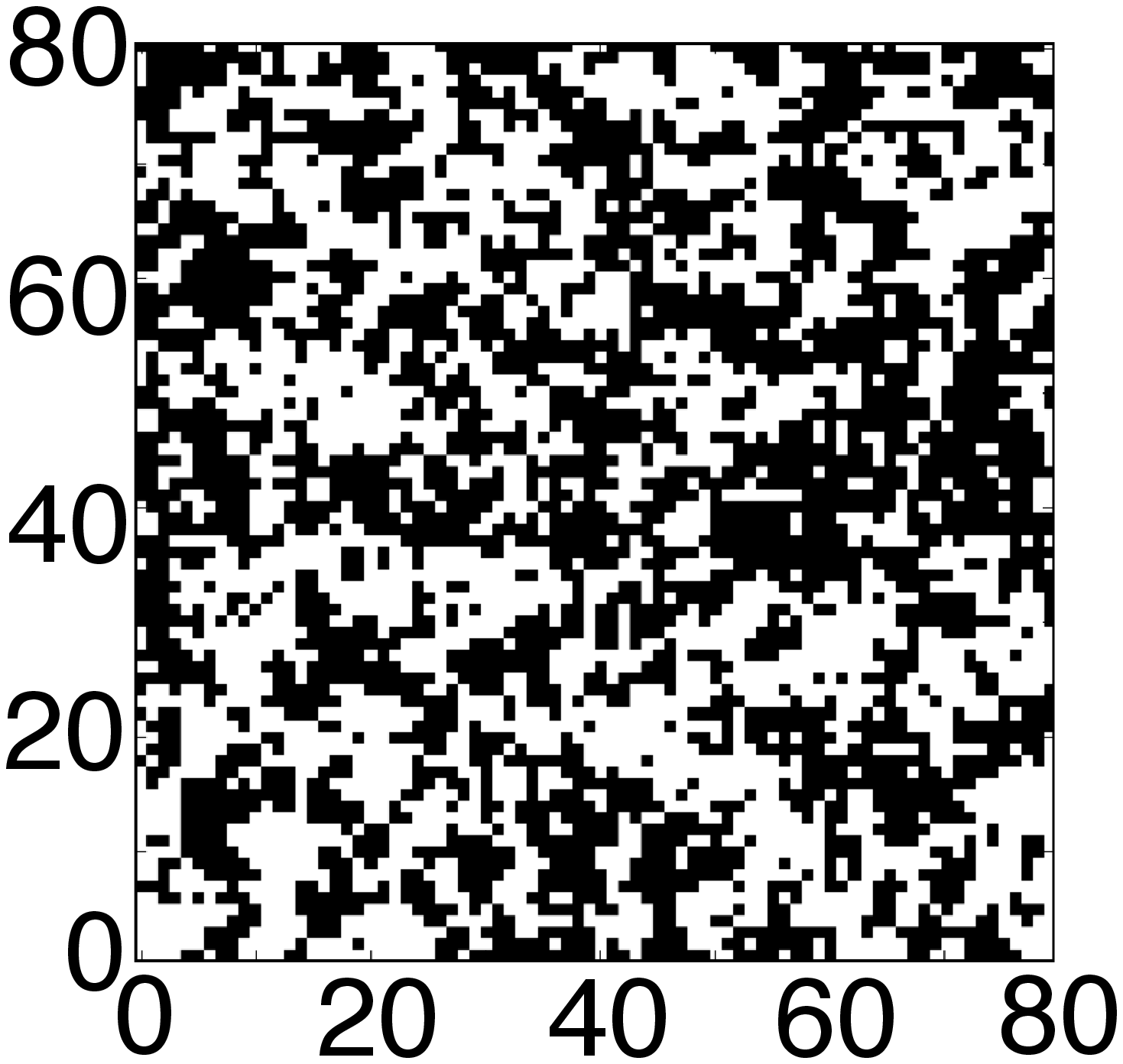}\label{fig:4sub1}}
\subfigure[]{\includegraphics[height=0.18\textwidth]{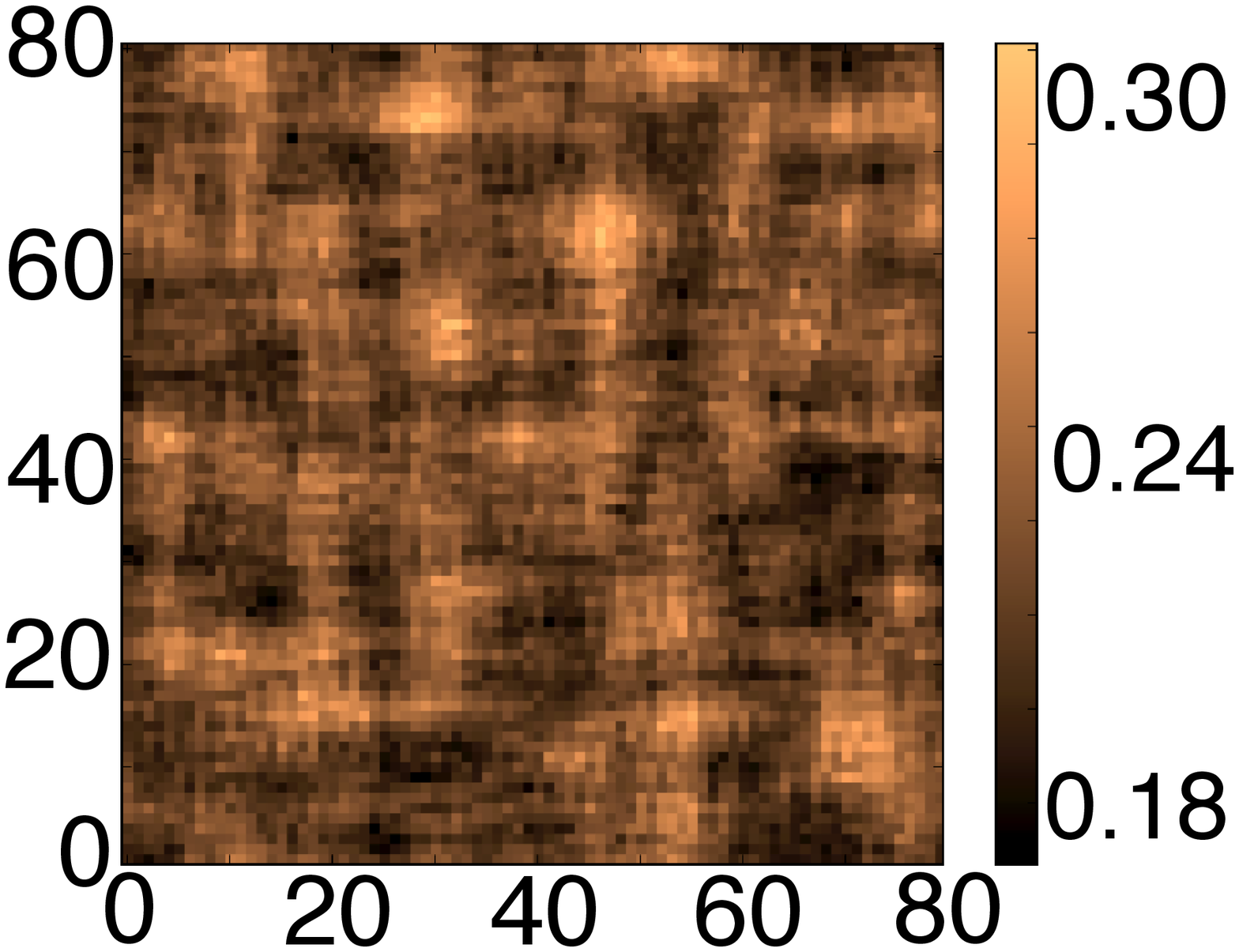}\label{fig:4sub2}}
\caption{{\bf Stability of the anisotropy under strong spatial disorder}. A typical off-critical ($T=1.4T_c$) configuration of a conserved order parameter Ising model is shown in (a), which represents a typical form of strong spatial disorder. The form of the LDOS (only the upper left quarter of the configuration is shown) in (b), at $\omega=t$ shows that the correspondence between the order parameter configuration and the form of the LDOS becomes very complicated. At very high energies ($\omega=t$), there is no evident correspondence between the positions of the domains and anisotropic features of the LDOS, due to the presence of a large collection of bound states generated by the jerky features of the order parameter, as discussed in Ref.~\cite{andersen06}}
\end{figure}

Given that off-diagonal disorder generates high energy bound states which distort the LDOS in complex ways~\cite{andersen06}, we studied the behavior of the LDOS in cases where the spatial form of the disorder has jerky features or where the amplitude has strong variations. When the defined domains have {\it jerky} features, as these would appear, for example in a conserved-order parameter Ising model near its critical point, we find that the LDOS high energy features have complex characteristics with no local anisotropy at regions where large domains exist (cf. Fig.~\ref{fig:4sub1},\ref{fig:4sub2}). However, amplitude (with no spatial jerkiness) variations of the off-diagonal disorder, as soon as they have zero global average, do not affect our qualitative conclusions. 

In conclusion we studied an explicit model of itinerant fermions which provides a simple explanation of the experimental facts observed in recent STM experiments. This model contains off-diagonal disorder, which is known to be ubiquitous in systems with tendency towards orthorhombic distortions (like \ybco and \lsco), and in other less distorted systems, such as \bscco and \ccoc, it may occur in the form of an inhomogeneous $\Delta_s$, modulated over distances of a few lattice sites. Off-diagonal disorder of this type does not distort the form of the LDOS at large distances and low energies, in contrast to potential impurities~\cite{balatsky03}. Furthermore its effect on the features of the LDOS, should not depend on the mechanism that drives them, which could be purely electronic as discussed in Ref.~\cite{vojta08}.

We would like to acknowledge rather helpful discussions with E.~A.~Kim, P.~Phillips, S.~Sachdev, J.~P.~Sethna. This research was partially supported by NSF DMR-0706379 (DG) and DOE-BES through DE-FG02-07ER46393  (SP).

\bibliography{offdiagonal.bib}

\begin{thebibliography}{10}
\bibitem{footnote-shortrange}%
  \BibitemOpen
  \bibinfo {note} {The size of the domains has been chosen to be $5a_0\times
  5a_0$, which is experimentally relevant. Variations around this size in one
  or both linear dimensions don't affect conclusions and neither do random
  spatial fluctuations in the magnitude of the disorder $\Delta^{(s)}$.}%
  \bibAnnoteFile{Stop}{footnote-shortrange}%
\end{thebibliography}

\end{document}